\documentclass[conference]{IEEEtran}
\ifCLASSINFOpdf
\else
\fi
\usepackage[cmex10]{amsmath}
\usepackage{amsfonts}
\usepackage{amssymb}
\usepackage{graphicx}

\newtheorem{theorem}{Theorem}

\newtheorem{definition}[theorem]{Definition}

\newcommand{\T}{T} 
\newcommand{\Y}[1]{\underline{Y}_{#1}}
\newcommand{\idx}{f} 		
\newcommand{\blk}{n}		

\begin{document}
%
\title{Interference Focusing for Simplified \\Optical Fiber Models with Dispersion}


\author{
\IEEEauthorblockN{Hassan Ghozlan}
\IEEEauthorblockA{Department of Electrical Engineering\\
University of Southern California\\
Los Angeles, CA 90089 USA\\
ghozlan@usc.edu}
\and
\IEEEauthorblockN{Gerhard Kramer}
\IEEEauthorblockA{
Institute for Communications Engineering \\
Technische Universit\"{a}t M\"{u}nchen \\
80333 M\"{u}nchen, Germany \\
gerhard.kramer@tum.de}
}


\maketitle

%
\IEEEpeerreviewmaketitle

\begin{abstract}
A discrete-time two-user interference channel model is developed 
that captures non-linear phenomena that arise in optical fiber communication 
employing wavelength-division multiplexing (WDM). 
The effect of non-linearity is that an amplitude variation on one carrier
induces a phase variation on the other carrier.
Moreover, the model captures the effect of group velocity mismatch
that introduces memory in the channel.
It is shown that both users can achieve the maximum pre-log factor of 1 simultaneously
by using an interference focusing technique introduced in an earlier work.
\end{abstract}

\section{Introduction}
\label{sec:intro}
The increase in traffic demand and the advances in optical technology  over the past two decades
made determining the capacity of optical fiber networks of great interest.
The optical fiber channel suffers from three main impairments of different nature: noise, dispersion, and Kerr non-linearity. 
The interaction between these three phenomena makes the problem of estimating the capacity challenging.
One approach is to solve the propagation equation numerically 
and then approximate the capacity by estimating the input and output statistics
(see, e.g., \cite{JLT2010} and references therein).
A second approach is to estimate the fiber capacity analytically, but then one must make simplifying assumptions.
In \cite{FOCUS2010}, we studied a simplified non-linear model in which $all$ orders of dispersion were neglected.
In this paper, we add first-order dispersion to the non-linear model,
which makes the model more realistic.

\section{Fiber Channel Models} 
\label{sec:fiber_channel}
Suppose that two optical fields at different carrier frequencies $\omega_1$ and $\omega_2$ are launched at the same location and propagate simultaneously inside the fiber. The fields interact with each other through the Kerr effect \cite[Ch. 7]{Agrawal}.
Specifically, neglecting fiber losses, the propagation is governed by the coupled non-linear Schr\"{o}dinger (NLS) equations \cite[p. 264, 274]{Agrawal}:
\begin{align}
	& i \frac{\partial A_1}{\partial z}  - \frac{\beta_{21}}{2} \frac{\partial^2 A_1}{\partial T^2} 
	+ \gamma_1 (|A_1|^2 + 2 |A_2|^2) A_1 = 0
	\label{eq:nls2_eq1} \\
	& i \frac{\partial A_2}{\partial z} - \frac{\beta_{22}}{2} \frac{\partial^2 A_2}{\partial T^2} 
	+ \gamma_2 (|A_2|^2 + 2 |A_1|^2) A_2 + i d \frac{\partial A_2}{\partial T} = 0
	\label{eq:nls2_eq2}
\end{align}
where $i = \sqrt{-1}$,
$A_k(z,T)$ is the time-retarded, slowly varying component of field $k$, $k=1,2$, the $\beta_{2k}$ are group velocity dispersion (GVD) coefficients, the $\gamma_k$ are non-linear parameters, and $d=\beta_{12}-\beta_{11}$ where the $\beta_{1k}$ are reciprocals of group velocities. The parameter $d$ is called group velocity mismatch (GVM).
To simplify the model, we assume that the second order dispersion effects are negligible, 
\textit{i.e.}, $\beta_{21}=\beta_{22}=0$.
In previous work \cite{FOCUS2010}, the group velocity mismatch was also ignored ($d = 0$),
which leads to a \textit{memoryless} channel model.
In this paper, we consider the case of non-zero GVM ($d \neq 0$),
which introduces memory as we shall see.
The coupled NLS equations \eqref{eq:nls2_eq1}--\eqref{eq:nls2_eq2} have the exact solutions \cite[p. 275]{Agrawal}
\begin{align}
	A_1(L,T) &= A_1(0,T) e^{i \phi_1(L,T)}, 
	\label{eq:gvm_exact1} \\	
	A_2(L,T) &= A_2(0,T-L d) e^{i \phi_2(L,T)},
	\label{eq:gvm_exact2}
\end{align}
where $L$ is the fiber length, and
the time-dependent non-linear phase shifts are obtained from
\begin{align}
	\phi_1(L,T) = \gamma_1 \Big( |A_1(0,T)|^2 L + 2 \int_{0}^L |A_2(0,T-z d)|^2 dz\Big), 
	\nonumber \\
	\phi_2(L,T) = \gamma_2 \Big( |A_2(0,T)|^2 L + 2 \int_{0}^L |A_1(0,T+z d)|^2 dz\Big).
	\nonumber
\end{align}
where $z=0$ is the point at which both fields are launched. 
Kerr non-linearity does not change the shape of the pulse in the time domain
but  causes {\it interference} through intensity-dependent phase shifts, which affects the spectrum of the pulse. 
The non-linear phase shift seen by a field due to the field itself is called self-phase modulation (SPM),
\textit{i.e.}, $\gamma_k L |A_k(0,T)|^2$ is an SPM term,
whereas the non-linear phase shift seen by a field due to other fields, which causes interference, 
is called cross-phase modulation (XPM). 
 
In \cite{FOCUS2010}, a discrete-time model was obtained by \textit{sampling} the fields at the transmitters and the receivers.
In general, this introduces noise with large variance at the receivers because wide-band noise is not removed by filtering. 
In the next section, we overcome this shortcoming by using \textit{matched filtering} before sampling at the receivers.
\section{Continuous-Time Model}
\label{sec:ct_model}
Consider the case of $d \neq 0$. 
Without loss of generality, suppose that $d>0$.
Let $\{x_k[m]\}_{m=0}^{\infty}$ be the data sequence sent by transmitter $k$.
Suppose that the transmitters employ square pulse shaping, \textit{i.e.}, 
the signal sent by transmitter $k$ is
\begin{align}
	A_k(0,\T) = \sum_{m=0}^{\infty} x_k[m] \ p(\T-m T_s).
	\label{eq:tx_k}
\end{align}
where
\begin{align}
 p(t) = \left\{ 
  \begin{array}{ll}
  \displaystyle \sqrt{E_s/T_s} ,	& 0 \leq t < T_s \\
  0,					& \text{otherwise}.
  \end{array}
 \right.
\label{eq:square_pulse_def}
\end{align}

The signal observed by receiver $k$ is
\begin{align}
	r_k(\T) = A_k(L,\T) + z_k(T), \quad k=1,2.
	\label{eq:rk}
\end{align}
where $z_k(T)$ is Gaussian white noise with 
$\mathbb{E}(z_k(T)) = 0$, and 
$\mathbb{E}(z_k(T) z_k^*(T+\tau)) = N \delta(\tau)$.
The processes $z_1(T)$ and $z_2(T)$ are independent.

The analysis of the two receivers is similar, hence we focus on receiver 1.
The received signal is fed to a bank of linear time-invariant (LTI) filters with impulse responses 
$\{h_{\idx}(\T)\}_{\idx \in \mathcal{F}_1}$,
where
\begin{align}
h_{\idx}(t) = p^*(-t) \exp( i 2\pi {\idx} t/T_s),
\end{align}
and $\mathcal{F}_1 \subset \mathbb{Z}=\{\ldots,-1,0,1,\ldots\}$. 
In other words, the bank of filters is a set of frequency-shifted matched filters whose impulse responses are orthogonal.
The choice of the set $\mathcal{F}_1$ is specified in section \ref{sec:ifocus}.

The output of the filter with index ${\idx}$ is
\begin{align*}
	y_{1,{\idx}}(\T) = r_1(\T) * h_{\idx}(\T), 
\end{align*}
where $*$ denotes convolution.
The ``noiseless part'' of the output of the filter with index ${\idx}$, $\tilde{y}_{1,{\idx}}(\T)$, is given by
\begin{align}
	\tilde{y}_{1,{\idx}}(\T) 
	&= (r_1(\T) - z_1(T)) * h_{\idx}(\T)  \\
	&= A_1(L,\T) * h_{\idx}(\T) 					 \\
	&= \left( A_1(0,\T) e^{i \phi_1(L,\T)} \right) * h_{\idx}(\T)  \\
	&= \sum_{m=0}^{\infty} x_1[m] \ \left( p(\T-m T_s) e^{i \phi_1(L,\T)} \right) * h_{\idx}(\T)  \\
	&= \sum_{m=0}^{\infty} x_1[m] \ \int_{- \infty}^{\infty} p(\tau-m T_s) p^*(\tau-\T)  \nonumber \\  
	& \qquad \qquad \qquad \qquad	  e^{i \phi_1(L,\tau)} e^{-i 2\pi {\idx} (\tau-\T)/T_s} d\tau.
\end{align}

By sampling the output signal $y_{1,{\idx}}(\T)$ at the time instants $\T = j Ts$, $j \in \{0,1,2,\ldots\}$, we have
\begin{align}
	\tilde{y}_{1,{\idx}}(j T_s) 
	&= \sum_{m=0}^{\infty} x_1[m] \ \int_{- \infty}^{\infty} p(\tau-m T_s) p^*(\tau-j T_s) \nonumber \\
	& \qquad \qquad \qquad \qquad	  e^{i \phi_1(L,\tau)-i 2\pi {\idx} (\tau-j T_s)/T_s} d\tau .
	\label{eq:noiseless_part_sampled}
\end{align}

Since $p(t)=\sqrt{E_s/T_s}$ for $t \in [0,T_s]$ and $p(t)=0$ for $t \notin [0,T_s]$, 
equation (\ref{eq:noiseless_part_sampled}) reduces to
\begin{align}
	\tilde{y}_{1,{\idx}}(j T_s) 
	&= x_1[j] \frac{E_s}{T_s} \ \int_{j T_s}^{j T_s+T_s} e^{i \phi_1(L,\tau)-i 2\pi {\idx} (\tau-j T_s)/T_s} d\tau.
	\label{eq:y1n_sampled}
\end{align}

We write $\phi_1(L,\tau)$ as
\begin{align}     
\phi_1(L,\tau) = \phi_{11}(L,\tau) + \phi_{12}(L,\tau),
\end{align}
in which we have defined 
\begin{align}
 \phi_{11}(L,t) &\stackrel{\Delta}{=}   \gamma_1  L \ |A_1(0,t)|^2,
 \label{eq:phi11_definition} \\
 \phi_{12}(L,t) &\stackrel{\Delta}{=} 2 \gamma_1 L_W \
 \frac{1}{T_s} \int_{t-L d}^{t} |A_2(0,\lambda)|^2 d\lambda,
 \label{eq:phi12_definition}
\end{align}
where $L_W \stackrel{\Delta}{=} T_s/d$. 
$L_W$ is called the walk-off distance.

Since $p(\lambda)=0$ for $\lambda \notin [0,T_s]$, we have
\begin{align}
 \phi_{12}(L,t) 
 &= \frac{2 \gamma_1 L_W}{T_s} \int_{t-L d}^{t} \sum_{m=0}^{\infty} |x_2[m]|^2 \ |p(\lambda-m T_s)|^2 d\lambda 	\\
 &= 2 \gamma_1 L_W \sum_{m=0}^{\infty} |x_2[m]|^2 \ \psi_{12}^{(m)}(t) ,
 \label{eq:psi12_expansion}
\end{align}
where $\psi_{12}^{(m)}(t)$ is defined as
\begin{align}
 \psi_{12}^{(m)}(t) \stackrel{\Delta}{=} \frac{1}{T_s} \int_{t-L d}^{t} |p(\lambda-m T_s)|^2 d\lambda.
 \label{eq:psi12m_definition}
\end{align}

Assume that $L d = M T_s$ for some $M \in \{1,2,3,\ldots\}$.
The larger the group velocity mismatch is, the larger $M$ is, the more memory the system has.
For $L d = M T_s$, we obtain
\begin{align}
  \psi_{12}^{(m)}(t) = 
  \frac{E_s}{T_s^2} \times
  \left\{
  \begin{array}{l}
   t-m T_s, 			  \quad \qquad	 m 		T_s \leq t < (m+1) 	 T_s	\\
   T_s,							\quad \qquad  (m+1) T_s \leq t < (m+M) 	 T_s	\\
   (m+M+1)T_s-t,\\	\quad \quad   (m+M) T_s \leq t < (m+M+1) T_s	\\
   0,								\quad \qquad \quad \text{otherwise.}
  \end{array}
  \right.
\label{eq:psi12m_Ld_equals_MTs}
\end{align}

For $ \tau \in [j T_s, j T_s+T_s]$, 
we have\footnote{We use the convention of setting the quantities that involve a negative time index to zero.}
\begin{align}
 \phi_{11}(L,\tau)
  &=   \gamma_1 L \frac{E_s}{T_s} |x_1[j]|^2 		,\\
 \phi_{12}(L,\tau) 
  &=   2 \gamma_1 L_W \frac{E_s}{T_s} \Big( 
  \sum_{r=1}^{M} |x_2[j-r]|^2  + 	\nonumber \\& \qquad
  ( |x_2[j]|^2 - |x_2[j-M]|^2) \frac{(\tau-j T_s)}{T_s}
 \Big) .
 \label{eq:phi11_and_phi12}
\end{align}

Hence, the integral in (\ref{eq:y1n_sampled}) evaluates to
\begin{align}
 \int_{j T_s}^{j T_s+T_s} {e^{ A + B {(\tau-j T_s)}/{T_s} }} d\tau
 	=  \left\{
 	\begin{array}{ll}
 	T_s \ e^A \frac{(e^B-1)}{B}, 	&	\text{if } B \neq 0	\\
 	T_s \ e^A,								&	\text{if } B   =  0 ,
  \end{array} 
  \right.
  \label{eq:integral_exp}
\end{align}
where
\begin{align}
 A   &= i h_{11} |x_1[j]|^2 + i h_{12} \sum_{r=1}^{M} |x_2[j-r]|^2, \\ 
 B   &= i h_{12}  ( |x_2[j]|^2 - |x_2[j-M]|^2 )	- 2 \pi {\idx}			\\
 h_{11} &= \gamma_1 L \frac{E_s}{T_s} \text{, and } h_{12} = 2 \gamma_1 L_W \frac{E_s}{T_s}.
\label{eq:rx1_a_and_b_in_terms_of_h_coeff}
\end{align}

Therefore, the ``noiseless part'' of the output of the filter with index ${\idx}$ at time $j$, $\tilde{y}_{1,{\idx}}[j]$, is given by
\begin{align}
&\tilde{y}_{1,{\idx}}[j]   = \label{eq:y1nj_without_noise} \\ & \ 
x_1[j] E_s \ \exp\Big( i h_{11} |x_1[j]|^2 + i h_{12} \sum_{r=1}^{M} |x_2[j-r]|^2 \Big) u_{1,{\idx}}[j],
\nonumber
\end{align}
where
\begin{align}
  u_{1,{\idx}}[j] 
  & = \left\{
  \begin{array}{ll}
	 \displaystyle
        \frac{\exp\left(i 2\pi(v_1[j] - {\idx})\right) - 1 }
	      {          			i 2\pi(v_1[j] - {\idx})            }	,&\text{ if } v_1[j] \neq {\idx}, \\
  			1																											,&\text{ otherwise,}
 \end{array}
 \right. 
 \label{eq:u1nj}
\end{align}
in which we have defined $v_1[j]$ as
\begin{align}
	v_1[j] \stackrel{\Delta}{=} h_{12}(|x_2[j]|^2 - |x_2[j-M]|^2)/ 2\pi.
	\label{eq:v1j}
\end{align}

The output of the filter with index ${\idx}$ at time $j$, $y_{1,{\idx}}[j]$, is
\begin{align}
	y_{1,{\idx}}[j] = y_{1,{\idx}}(j T_s) = \tilde{y}_{1,{\idx}}[j] + z_{1,{\idx}}[j],
	\label{eq:y1nj}
\end{align}
where 
$z_{1,{\idx}}[j] = z_1(T) * h_{\idx}(T) |_{T=j T_s}$.
The variable $z_{1,{\idx}}[j]$ is Gaussian with mean 0 and variance $N E_s$.
Moreoever, due to the orthogonality of the filter bank impluse responses, we have
$\mathbb{E}(z_{1,{\idx}}[j] z_{1,\tilde{\idx}}^*[j]) = 0$ for all ${\idx} \neq \tilde{\idx}$,
which implies, because of Gaussianity, that 
the random variables $\{z_{1,{\idx}}[j]\}_{\idx \in \mathcal{F}_1}$ are independent.

\section{Discrete-Time Model}
\label{sec:dt_model}
Transmitter $k$ sends a string $X_k^{\blk} = (X_{k}[1], X_{k}[2], \cdots, X_{k}[{\blk}])$ of symbols 
while receiver $k$ observes $\Y{k}^{\blk} = (\Y{k}[1], \Y{k}[2], \cdots, \Y{k}[{\blk}])$.
The input $X_k[j]$ of transmitter $k$ to the channel at time $j$ is a scalar,
whereas the channel output $\Y{k}[j]$ at receiver $k$ at time $j$ is a vector
whose components are $Y_{k,{\idx}}[j]$, $f \in \mathcal{F}_k$.
The input-output relations are:
\begin{align}
Y_{1,{\idx}}[j] &= X_1[j] \ \exp\Big( i h_{11} |X_1[j]|^2 + \nonumber \\ & \qquad  
																 i h_{12} \sum_{r=1}^{M} |X_2[j-r]|^2 \Big) U_{1,{\idx}}[j] + Z_{1,{\idx}}[j],	
\\
Y_{2,{\idx}}[j] &= X_2[j] \ \exp\Big( i h_{21} \sum_{r=1}^{M} |X_1[j+2M-r]|^2 + \nonumber \\ & \qquad
						  									 i h_{22} |X_2[j+M]|^2 \Big) U_{2,{\idx}}[j] + Z_{2,{\idx}}[j],
\label{eq:model_y1nj_y2nj}
\end{align}
where 
\begin{align}
  U_{k,{\idx}}[j] 
  & = \left\{
  \begin{array}{ll}
	 \displaystyle
        \frac{\exp\left(i 2\pi(V_k[j] - {\idx})\right) - 1 }
	      {          			i 2\pi(V_k[j] - {\idx})            }	,&\text{ if } V_k[j] \neq {\idx}, \\
  			1																											,&\text{ otherwise,}
 \end{array}
 \right. 
 \label{eq:model_uknj}
\end{align}
in which we have defined $V_k[j]$ as
\begin{align}
	V_1[j] &\stackrel{\Delta}{=} h_{12}(|X_2[j]		|^2 - |X_2[j-M]|^2)/ 2\pi, \\
	V_2[j] &\stackrel{\Delta}{=} h_{21}(|X_1[j+2M]|^2 - |X_2[j+M]|^2)/ 2\pi.
	\label{eq:model_vk}
\end{align}
$Z_{k,{\idx}}[j]$ models the noise at receiver $k$ at time $j$, 
the random variables $\{Z_{k,{\idx}}[j]\}_{k,\idx,j}$ are independent circularly-symmetric complex Gaussian random variables with mean 0  and variance $N$.
We regard the $h_{k\ell}$ as \textit{channel coefficients} that are time invariant and known globally.
The following power constraint is imposed:
\begin{align} \label{eq:power-constraint}
	\frac{1}{{\blk}} \sum_{j=1}^{{\blk}} \mathbb{E}\left[ |X_{k}[j]|^2 \right] \le P_k, \quad k=1,2.
\end{align}

A \textit{scheme} is a collection $\{ (\mathcal{C}_1(P_1,P_2,N),\mathcal{C}_2(P_1,P_2,N) )\}$ of pairs of codes
such that at $(P_1,P_2,N)$, user $k$ uses the code $\mathcal{C}_k(P_1,P_2,N)$ 
that satisfies the power constraint and achieves an information rate $R_k(P_1,P_2,N)$ where $k=1,2$.
We make a distiction betweeen two limiting cases:
1) fixed noise with growing powers, and 
2) fixed powers with vanishing noise.

\begin{definition}
The \textit{high-power} pre-log pair $(\overline{r}_1, \overline{r}_2)$ is achieved
by a scheme $\{ (\mathcal{C}_1(P_1,P_2,N),\mathcal{C}_2(P_1,P_2,N) )\}$ if
the rates satisfy
\begin{equation}
\overline{r}_k(\beta,N) = \lim_{P_k \rightarrow \infty: P_2 = P_1^\beta} \frac{R_k(P_1,P_2,N)}{\log(P_k/N)}
\text{ for } k=1,2,
\label{eq:high-power_prelog_userk_def}
\end{equation}
where $\beta$ is real.
\end{definition}
\begin{definition}
The \textit{low-noise} pre-log pair $(\underline{r}_1, \underline{r}_2)$ is achieved
by a scheme $\{ (\mathcal{C}_1(P_1,P_2,N),\mathcal{C}_2(P_1,P_2,N) )\}$ if
the rates satisfy
\begin{equation}
\underline{r}_k(P_1,P_2) = \lim_{N \rightarrow 0} \frac{R_k(P_1,P_2,N)}{\log(P_k/N)}
\text{ for } k=1,2.
\label{eq:low-noise_prelog_userk_def}
\end{equation}
\end{definition}


The (high-power or low-noise) pre-log pair $(1/2,1/2)$ can be achieved if both users use only amplitude modulation. 
In this work, we show that the high-power pre-log pair $(1,1)$ can be achieved 
for any real $\beta$ and positive $N$ through \textit{interference focusing},
which was introduced in \cite{FOCUS2010}.
Determining the maximal achievable low-noise pre-log pair is a subject of ongoing work.
\section{Interference Focusing}
\label{sec:ifocus}
We use \textit{interferencing focusing}, \textit{i.e.}, focus the phase interference on \textit{one} point 
by imposing the following constraints on the transmitted symbols:
\begin{align}
 h_{21} |X_1[j]|^2 &= 2\pi \tilde{n}_1, \ \tilde{n}_1 = 1,2,3,\ldots\\
 h_{12} |X_2[j]|^2 &= 2\pi \tilde{n}_2, \ \tilde{n}_2 = 1,2,3,\ldots
\end{align}
which ensures that the XPM interference is eliminated.

Because of other constraints, e.g., power constraint, 
only a subset of the allowed \textit{rings} is actually used, \textit{i.e.}, we choose
\begin{align}
 |X_1[j]|^2 \in \mathcal{P}_1 &= \{ 2\pi \tilde{n}_1 / h_{21} , \ \tilde{n}_1 \in \mathcal{N}_1 \}\\
 |X_2[j]|^2 \in \mathcal{P}_2 &= \{ 2\pi \tilde{n}_2 / h_{12} , \ \tilde{n}_2 \in \mathcal{N}_2 \}
\end{align}
where $\mathcal{N}_1,\mathcal{N}_2 \subset \mathbb{N}=\{1,2,3,\ldots\}$.
In this case, $V_k[j] \in \mathcal{V}_k$, for $k=1,2$, where 
\begin{align}
	\mathcal{V}_k \stackrel{\Delta}{=} \left\{ m_1 - m_2: m_1 \in \mathcal{N}_{3-k}, m_2 \in \mathcal{N}_{3-k} \right\},
\end{align}
which leads us to choose the sets of ``normalized frequencies'' $\mathcal{F}_1$ and $\mathcal{F}_2$ 
of the filter banks at the receivers as
$\mathcal{F}_1 = \mathcal{V}_1$ and $\mathcal{F}_2 = \mathcal{V}_2$.

Thus, under interference focusing, the output at receiver 1 is a vector $\underline{Y}_1$, 
whose components are $\{Y_{1,{\idx}}\}_{{\idx} \in \mathcal{V}_1}$,
where\footnote{We drop the time index for notational simplicity.}
\begin{align}
Y_{1,{\idx}}   = X_1 \ \exp\left( i h_{11} |X_1|^2 \right) U_{1,{\idx}} + Z_{1,{\idx}},
\label{eq:y1n}
\end{align}
and where 
\begin{align}
  U_{1,{\idx}}
  = \left\{ \begin{array}{ll}
		1	,&\text{ if } V_1 = {\idx}, \\
  	0	,&\text{ otherwise.}
 \end{array} \right.
 \label{eq:u_1n_ifocus}
\end{align}

This means that, under interference focusing, 
exactly one filter output among all the filters ``contains the signal'' corrupted by noise,
while all other filters put out noise.

We wish to compute the information rate $R_1 = I(X_1;\Y{1})$.
From the chain rule, and non-negativity of mutual information,
we have
\begin{align}
	R_1 = I(X_1;\Y{1}) 
	&\geq I\left(X_1; Y_{1,V_1} \right) \\
	&= I\left( X_1;X_1 e^{i h_{11}|X_1|^2}+Z_{1,V_1} \right).
\end{align}

It was shown in \cite{FOCUS2010} that using interference focusing,
we have
\begin{align}
	\lim_{P_1 \rightarrow \infty} \frac{I(X_1;X_1 e^{i h_{11}|X_1|^2}+Z_{1,V_1})}{\log(P_1)}
	\geq 1,
\end{align}
which implies that $\overline{r}_1 \geq 1$.

Therefore, the (high-power) pre-log $\overline{r}_1=1$ is achievable.
Similarly, one can show that $\overline{r}_2=1$ is also achievable, and hence,
the pre-log pair $(1,1)$ is achievable.
Using a simple genie-aided argument\footnote{The genie gives each receiver the message it is not interested in.}, 
we find that the maximal pre-log pair is $(1,1)$.

The following toy example illustrates our receiver structure, 
and the role interference focusing plays in choosing its parameters.

\begin{figure}
	\centering
		\includegraphics[width=0.24\textwidth]{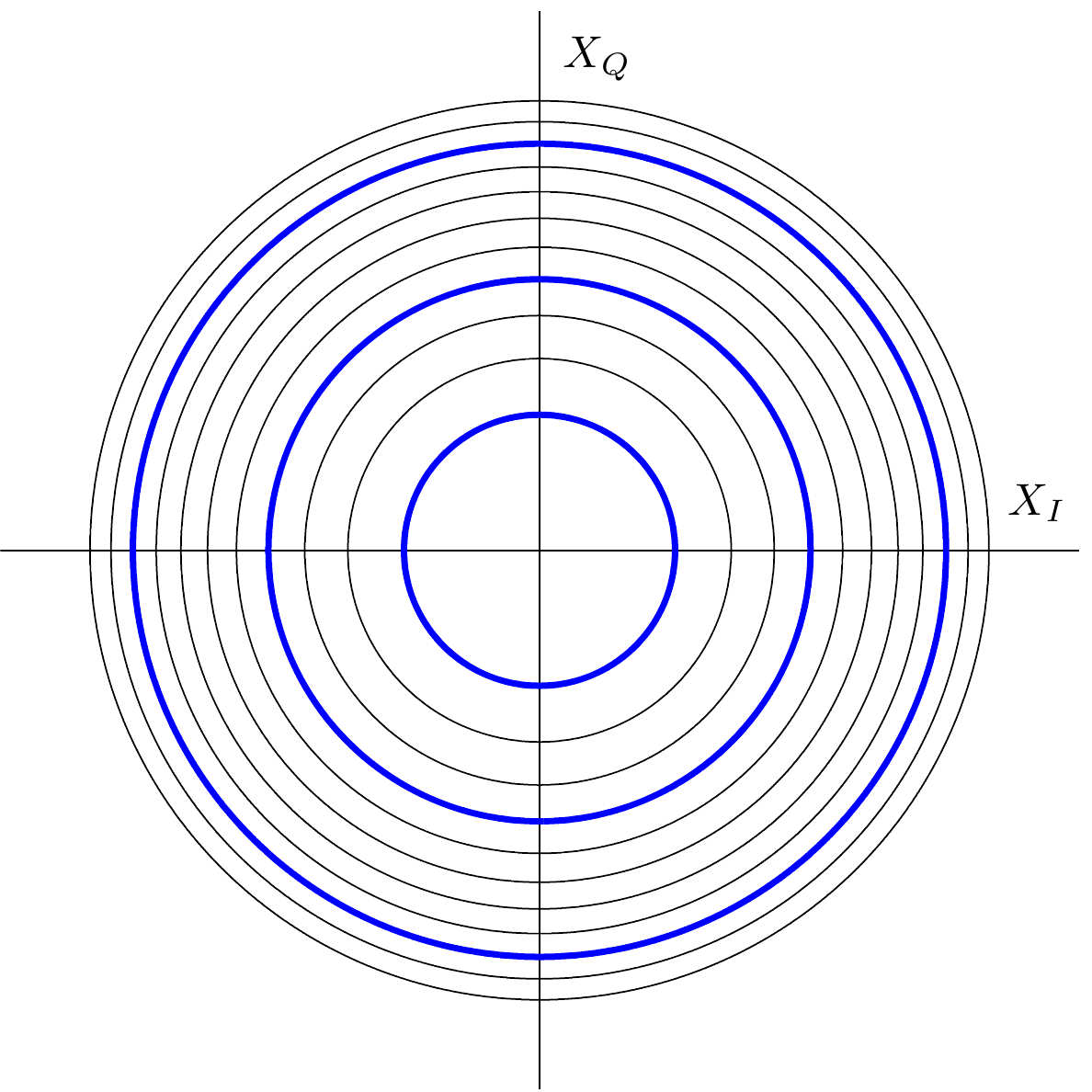}
		\includegraphics[width=0.24\textwidth]{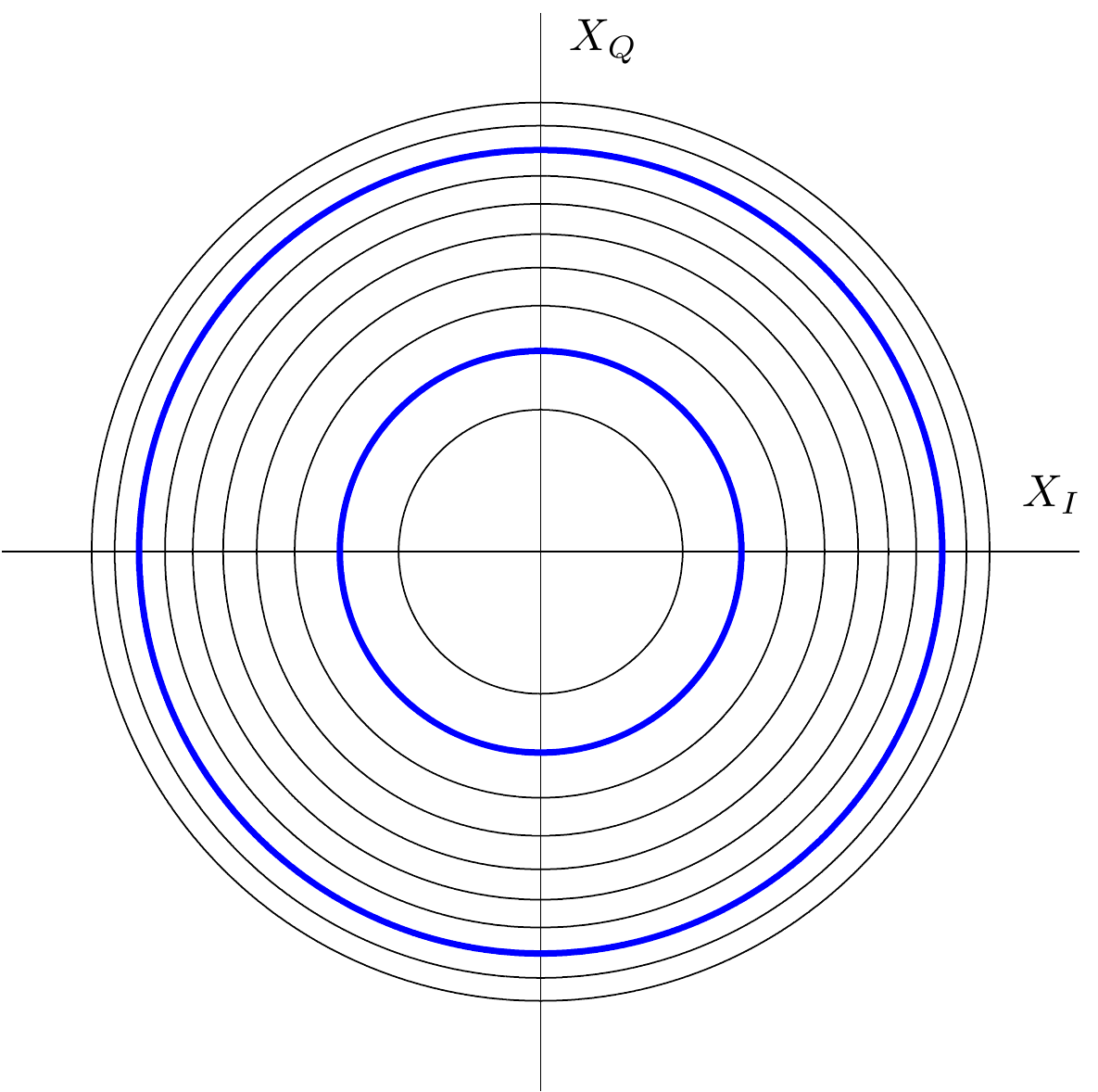}
	\caption{Ring modulation used by transmitter 1 (left) and transmitter 2 (right).
	The thin lines are the rings allowed by interference focusing, 
	and the thick blue lines are the rings selected for transmission.}
	\label{fig:constell}
\end{figure}
\begin{figure}
	\centering
		\includegraphics[width=0.5\textwidth]{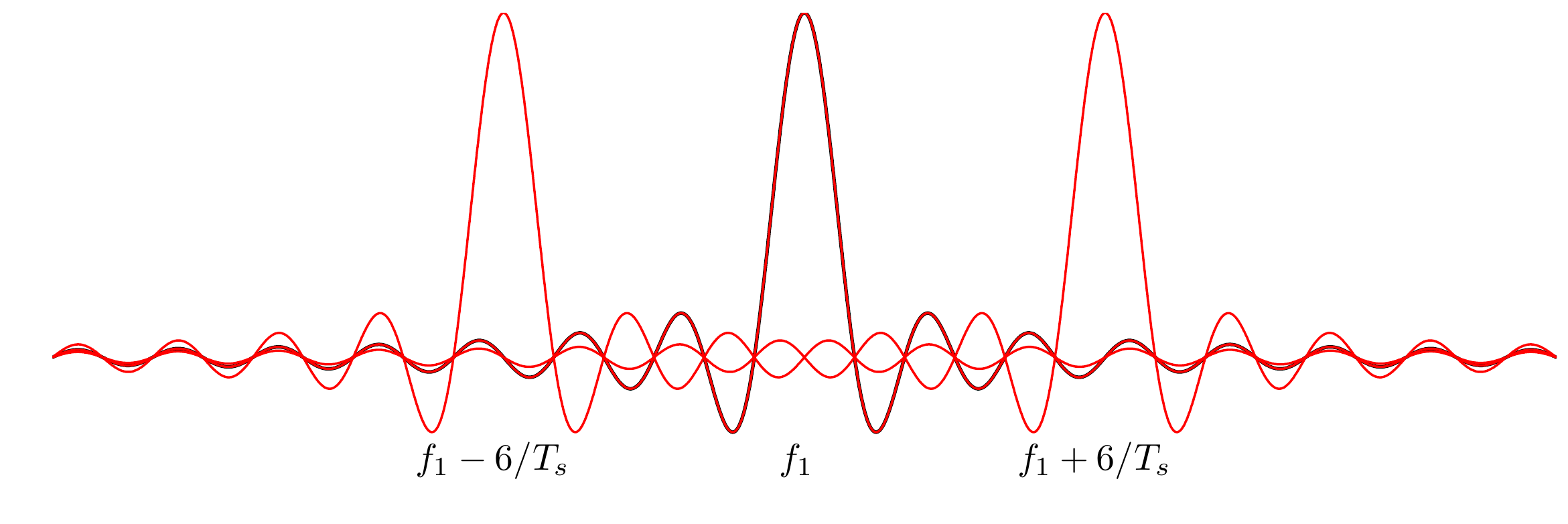} \\
		\includegraphics[width=0.5\textwidth]{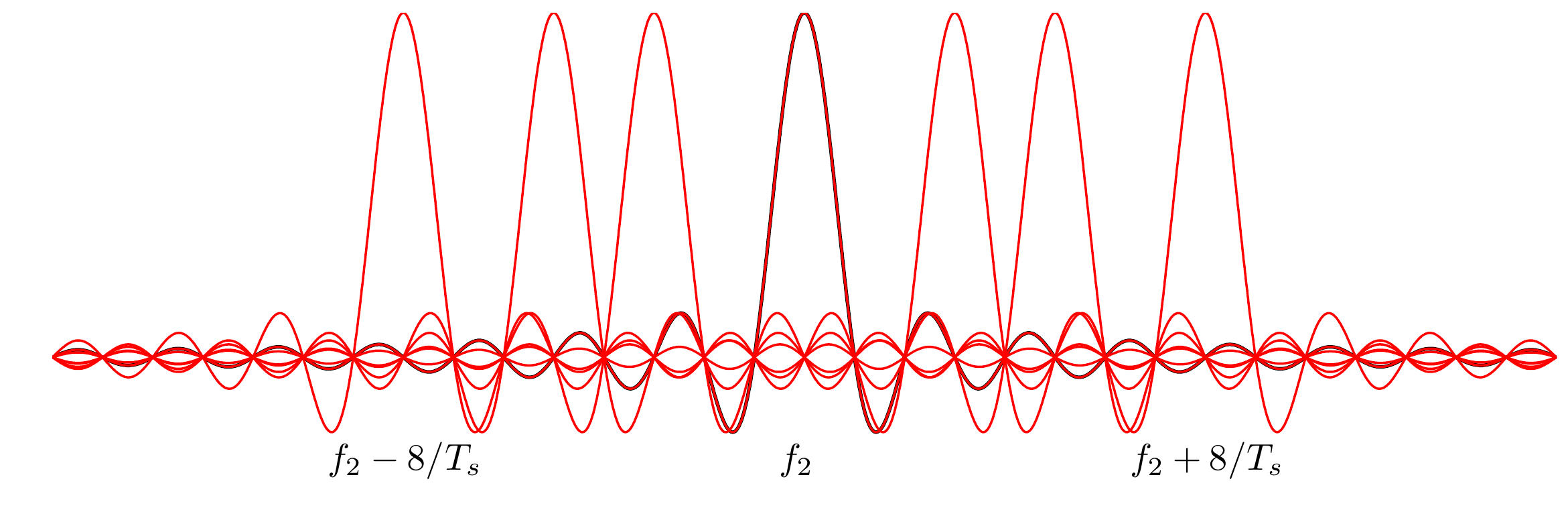}
	\caption{Frequency responses of the filters at receivers 1 (top) and 2 (bottom).}
	\label{fig:filter_bank}
\end{figure}

\textit{Example:}
Suppose that
$h_{12} = 5$, 
$h_{21} = 4$, 
$P_1 = 8$, 
and
$P_2 = 7$.
Suppose that the users choose the sets of rings
$\mathcal{N}_1=\{1,4,9\}$ and
$\mathcal{N}_2=\{2,8\}$ (see Fig. \ref{fig:constell}), 
\textit{i.e.},
the power levels are
$\mathcal{P}_1 = \{0.5\pi, 2\pi, 4.5\pi \}$
and
$\mathcal{P}_2 = \{0.8\pi, 3.2\pi \}$.
These choices satisfy the power contraints 
and eliminate the interference.
The parameters of the filter bank are 
$\mathcal{F}_1 = \mathcal{V}_1 = \{-6,0,6\}$ and
$\mathcal{F}_2 = \mathcal{V}_2 = \{-8,-5,-3,0,3,5,8\}$.
In other words, receiver 1 has 3 filters
whose frequency responses are sinc functions
centered at $f_1-6/T_s$, $f_1$, and $f_1+6/T_s$,
whereas
receiver 2 has 7 filters
whose frequency responses are sinc functions
centered at 7 different frequencies 
(see Fig. \ref{fig:filter_bank}).
This shows that, because of the non-linearity, 
the receivers need to extract information from a ``bandwdith'' larger than
the ``bandwidth'' of the transmitted signal.

\section{Conclusion}
We introduced a discrete-time two-user interference channel model 
based on a simplified optical fiber model. 
We assumed that second order dispersion is negligible.
However, we considered non-zero group velocity mismatch 
as well as non-linearity. 
We showed that our discrete-time model is justified 
by using square pulse shaping at the transmitters
and a bank of frequency-shifted matched filters at the receivers.
We proved that both users can achieve a high-power pre-log of 1 simultaneously
by using interference focusing, 
thus exploiting all the available amplitude and phase degrees of freedom.


\section*{Acknowledgment}
H. Ghozlan was supported by a USC Annenberg Fellowship and NSF Grant CCF-09-05235.
G. Kramer was supported by an Alexander von Humboldt Professorship endowed by
the German Federal Ministry of Education and Research.



%


\bibliographystyle{unsrt}
\bibliography{optic_ref6}

\end{document}